\documentclass[runningheads]{llncs}

\usepackage[utf8]{inputenc}
\usepackage{cite}
\usepackage[english]{babel}
\usepackage{authblk}
\usepackage{graphicx}
\usepackage{amsmath}
\usepackage{float}
\usepackage{hyperref}
\usepackage{multirow}
\usepackage{setspace}
\usepackage{enumerate}
\usepackage[labelfont=bf]{caption}

\usepackage{graphicx}
\begin{document}
\title{Probabilistic Multivariate Early Warning Signals}
%
%\titlerunning{Abbreviated paper title}
% If the paper title is too long for the running head, you can set
% an abbreviated paper title here
%
\author{Ville Laitinen\inst{1}\orcidID{0000-0002-2653-1673} \and
Leo Lahti\inst{1}\orcidID{0000-0001-5537-637X}}
\authorrunning{V. Laitinen and L. Lahti}
% First names are abbreviated in the running head.
% If there are more than two authors, 'et al.' is used.
%
\institute{Department of Computing, University of Turku, Turku, Finland\\\email{\{ville.laitinen, leo.lahti\}@utu.fi}
%\and
%Springer Heidelberg, Tiergartenstr. 17, 69121 Heidelberg, Germany
%\email{lncs@springer.com}\\
%\url{http://www.springer.com/gp/computer-science/lncs} \and
%ABC Institute, Rupert-Karls-University Heidelberg, Heidelberg, Germany\\
%\email{\{abc,lncs\}@uni-heidelberg.de}
}
\maketitle              % typeset the header of the contribution

%%%%%%%%%%%%%%%%%%%%%%%%%%%%%%%%%%%%%%%%%%%%%%%%%%%%%%%%
%%%%%%%%%%%%%%%%%%%%%%% ABSTRACT %%%%%%%%%%%%%%%%%%%%%%%
%%%%%%%%%%%%%%%%%%%%%%%%%%%%%%%%%%%%%%%%%%%%%%%%%%%%%%%%

\begin{abstract}
% The abstract should briefly summarize the contents of the paper in 15--250 words.

A broad range of natural and social systems from human microbiome to financial markets can go through critical transitions, where the system suddenly collapses to another stable configuration. Critical transitions can be unexpected, with potentially catastrophic consequences. Anticipating them early and accurately can facilitate controlled system manipulation and mitigation of undesired outcomes. Obtaining reliable predictions have been difficult, however, as often only a small fraction of the relevant variables can be monitored, and even minor perturbations can induce drastic changes in fragile states of a complex system. Data-driven indicators have been  proposed as an alternative to prediction and signal an increasing risk of forthcoming transitions. Autocorrelation and variance are examples of generic indicators that tend to increase at the vicinity of an approaching tipping point across a range of systems. An important shortcoming in these and other widely studied indicators is that they deal with simplified one-dimensional representations of complex systems. Here, we demonstrate that a probabilistic data aggregation strategy can provide new ways to improve early warning detection by more efficiently utilizing the available information from multivariate time series. In particular, we consider a probabilistic variant of a vector autoregression model as a novel early warning indicator and argue that it has theoretical advantages related to model regularization, treatment of uncertainties, and parameter interpretation. We evaluate the performance against alternatives in a simulation benchmark and show improved sensitivity in EWS detection in a common ecological model encompassing multiple interacting species.

\end{abstract}

%%%%%%%%%%%%%%%%%%%%%%%%%%%%%%%%%%%%%%%%%%%%%%%%%%%%%%%%
%%%%%%%%%%%%%%%%%%%%%%% INTRODUCTION %%%%%%%%%%%%%%%%%%%
%%%%%%%%%%%%%%%%%%%%%%%%%%%%%%%%%%%%%%%%%%%%%%%%%%%%%%%%

%\newpage

\section{Introduction}

The ability to anticipate and manage change plays a critical role in diverse domains, ranging from biomedicine to ecology, economics, or climate change \cite{Scheffer2009book}. Natural and social systems are inherently complex arrangements of smaller units and their interactions. Despite their complexity and size, such systems often exhibit remarkable stability where perturbations have only minor, and often only temporary and reversible effects on the system. However, when the conditions are stretched far enough, a system may pass a critical threshold, a tipping point, leading to a rapid and potentially irreversible reorganization. Such phenomena can be observed across many different scenarios \cite{lenton_review_2020}, including ecosystems \cite{Scheffer2001, TippingElements}, epidemics \cite{covid_EWS}, and climate \cite{climate_lenton}. As large transitions may have far-reaching consequences, the ability to anticipate them can provide valuable tools to manage change. 

Generic early warning signals (EWS) have been introduced to detect signs that could alarm us about approaching tipping points in complex systems  \cite{Scheffer_EWS, clements_review_2018}. The main challenge for such analysis is that the sequence of events leading to critical transitions can be subtle and gradual, with little or no apparent changes in the observable system state \cite{Hastings_no_warning_2010}. However, the underlying system dynamics may change in ways that can be observed and quantified. For instance, a decreasing resilience after perturbations \cite{Scheffer_EWS}, or {\it critical slowing down}, is often associated with an approaching tipping point. Increasing autocorrelation, variance, and other statistical properties can indicate critical slowing down in the vicinity of a tipping point, and they have become some of the most robust and widely utilized EWS \cite{Dakos_2012, Clements_2015, dakos_robustness_of_variance_and_autocorrelation}. An important property of such statistical indicators is their generality. They provide data-driven quantification that require minimal understanding of the data generating processes. This facilitates the detection of early warnings even when accurate mechanistic modeling is infeasible due to the complexity of the phenomena and limitations in data collection, and makes the generic EWS indicators applicable across a broad range of different systems. 

Despite the recent advances, the generic EWS often rely on the availability of sufficiently long and dense time series and manual parameter tuning. The ability to efficiently utilize data and detect EWS from more limited time series would be important in many application fields such as ecology and human medical studies where the sample sizes can be remarkably low due to ethical, financial, or other constraints. Quantification of uncertainty is another key aspect in EWS analysis. Data is always limited, and may come with uneven observation times, measurement noise, or possible biases. The ability to quantify and control uncertainty is particularly relevant with limited sample sizes. The probabilistic framework provides tools to incorporate uncertainty and prior information into the models \cite{gelman_BDA}, potentially leading to a more sensitive EWS detection. We recently demonstrated this by introducing a univariate probabilistic method for EWS detection \cite{pTVAR}, showing improvements in automated model selection and increased sensitivity in EWS detection performance. 

An important limitation in this and other previously proposed EWS indicators is their emphasis on univariate representations. This provides convenient, intuitively appealing, and robust ways to summarize changes in complex systems. Yet, the reliance on one-dimensional summaries may neglect potentially relevant information that could enhance the EWS detection. An enhanced use of multivariate observations in EWS design could provide improved sensitivity especially in shorter multivariate time series. The advantages of probabilistic methods in treating uncertainties, and the potential for improving EWS detection with data aggregation strategies motivated us to investigate the possibility of extending our earlier work on probabilistic EWS into the multivariate domain. 

In this work we design and investigate a novel probabilistic EWS indicator for multivariate systems. More specifically, we formulated and implemented a probabilistic variant of the time-varying vector autoregressive-1 model. A non-probabilistic version of a similar model was recently studied in \cite{Ives_Dakos_tvarss}. The probabilistic version allows alternative ways to pool information across the multivariate time series and deal with uncertainties in the modeling process. Moreover, it supports automated parameter inference, helping to circumvent the need to manually select model parameters, such as sliding window size, which have posed problems in many EWS methods. Besides these and other theoretically appealing properties, simulations based on a well-studied ecological model demonstrate good overall performance against the currently available alternatives. 

The work is structured as follows. In \emph{Methods} we describe the novel approach along with a set of previously studied EWS indicators. We then compare these indicators in a simulation study in \emph{Results} and, finally, conclude in \emph{Discussion} with some directions for further extension.

%\emph{Critical slowing down} refers to the tendency of the system dynamics to slow down near a tipping point. 

%This has provided early warning indicators that detect resilience loss in one dimensional time series.

%%%%%%%%%%%%%%%%%%%%%%%%%%%%%%%%%%%%%%%%%%%%%%%%%%%%%%%%
%%%%%%%%%%%%%%%%%%%%%%% METHODS %%%%%%%%%%%%%%%%%%%%%%%%
%%%%%%%%%%%%%%%%%%%%%%%%%%%%%%%%%%%%%%%%%%%%%%%%%%%%%%%%

\section{Methods}

In this section, we provide a short overview of the currently available, related methods based on a recent review \cite{weinans_2021}. We formulate the probabilistic variant, tvPVAR(1), and describe the simulation model that is used to generate data for the experiments.

% Variance based indicators:
% 1st eigenvalue,  PC1 variance, Average variance,  Maximum covariance,  Explained variance, Maximum absolute cross-correlation, Maximum variance,

%%%%%%%%%%%%%%%%%%%%%%%%%%%%%%%%%%%%%%%%%%%%%%%%%%%%%%%%%%%%%%%%%%%%%%%%%%%%%%%%%%%%%%%%%%
\subsection{Autocorrelation based EWS}
%%%%%%%%%%%%%%%%%%%%%%%%%%%%%%%%%%%%%%%%%%%%%%%%%%%%%%%%%%%%%%%%%%%%%%%%%%%%%%%%%%%%%%%%%%

Let us start by summarizing relevant methodology based a recent comparison between currently available indicators for detecting EWS in multivariate data \cite{weinans_2021}. In the present work we focus on autocorrelation-based indicators since these have shown robustness compared to the alternatives \cite{dakos_robustness_of_variance_and_autocorrelation, weinans_2021}, and can be naturally extended into the probabilistic framework that we explore in this study. More specifically, the methods detecting changes in lag-1 autocorrelation are based on the standard autoregressive-1 process, AR(1), defined by the recursion

\begin{equation}
\label{eqn:ar1_definition}
	X_{t+1} = \phi X_{t} + \sigma \epsilon_t
\end{equation} 

where \(X_t\) is the state variable at time \(t\), \(\phi\) the autoregressive parameter, \(\epsilon_t\) a zero-mean, unit variance Gaussian random variable scaled with \(\sigma\). The main interest here lies in the autoregressive parameter \(\phi\), which directly measures the lag-1 autocorrelation of the system.

Many variants and extensions of the AR(1) process have been studied in univariate context \cite{Dakos_2012}, and applications in multivariate data are also possible. For instance, \emph{maximum autocorrelation} (ac/max) \cite{dakos_2018_multispecies} is based on fitting the AR(1) model separately to each node, or feature, of the system and then selecting the one with the highest autocorrelation as a proxy for the entire system. Other options include \emph{average autocorrelation} (ac/mean) across the features, or \emph{degenerate fingerprinting} which measures autocorrelation along the first principal component of the multivariate data \cite{degen_finger}. Min/Max autocorrelation factors analysis (MAF) \cite{MAF} is another method based on dimension reduction that aims to identify the subspace with the highest autocorrelation in a multidimensional system. This algorithm generates a set of vectors (MAFs) to project the multidimensional data onto a subspace where autocorrelation is maximized. Eigenvalues of the MAF subspace quantify autocorrelation in the respective directions, with lower values indicating higher autocorrelation. In addition to \emph{MAF eigenvalues} (eigen/MAF), we used autocorrelation (ac/MAF) and variance (var/MAF) projected onto the 1st MAF as indicators. For a more detailed description of the these methods, see \cite{MAF}. We did omit some of the methods considered in  \cite{weinans_2021}, such as information dissipation length \cite{IDL} and time \cite{IDT}, since they require larger amounts of data, and our interest lies mainly in practically motivated situations where the sample sizes are modest.

The EWS detection based on these previously suggested indicators was carried out following standard procedures \cite{Dakos_2012}. We estimated the early warning indicators in sliding windows along the time series, resulting in a trajectory of the indicator, which is the autoregressive parameter \(\phi\) in our case. Except where otherwise noted, we set the sliding window to 50\% of length of the time series, which is a common default choice in the EWS literature. In order to remove the effect of non-stationary trends in the data that could lead to spurious conclusions \cite{Dakos_2012}, we used Gaussian detrending (R function \emph{stats::smooth}) as a preprocessing step before quantifying the indicator. We used a bandwidth of 10\% of the total time series length, except where otherwise noted, which we chose based on visual assessment; the bandwidth length was chosen so that it removes long-term mean level variations unrelated to the short term correlation structure while aiming to avoid overfitting to short-term variations. 

We then measured the strength of each estimated EWS by computing Kendall's rank correlation \(\tau\) between the estimated autocorrelation trajectory and the autoregressive parameter \(\phi\). The rank correlation receives values in [-1, 1], with \(\tau = 1\) indicating a monotonously increasing trajectory. The rank correlation is defined as \(\tau = (N_{\text{concordant pairs}} - N_{\text{disconcordant pairs}})/N_{\text{all pairs}}\), where \(N\) refers to the number of elements in the subscript set, and a pair \( {(t_i, \phi_i), (t_j, \phi_j)}\) is said to be concordant if \(t_i \leq t_j\) implies \(\phi_i \leq \phi_j\) and disconcordant otherwise. 

For hypothesis testing on these EWS indicators, we utilized the so-called surrogate data analysis methods \cite{Dakos_2012}. This technique generates an approximate sampling distribution for Kendall's \(\tau\), which is then compared to the actual point estimate. The sampling distribution represents results that would be recovered under the null hypothesis that the indicator trend has arisen simply by change. We generated a collection of time series from the simulation model presented below in the subsection \ref{simulation_model}. We used constant parameters that produce data where the conditions remain constant and any estimated parameter trajectory is expected to have no correlation with time (\(\tau = 0\)). We generated 500 replicates of surrogate data for each experimental condition, and then estimated the EWS indicators and Kendall's rank correlations for these surrogate data sets, yielding approximate sampling distributions under the null hypothesis. \emph{P}-values for a positive trend value were then computed as the proportion of the sampling distribution that exceeded (or were identical to) the actual point estimate. 

%%%%%%%%%%%%%%%%%%%%%%%%%%%%%%%%%%%%%%%%%%%%%%%%%%%%%%%%%%%%%%%%%%%%%%%%%%%%%%%%%%%%%%%%%%
\subsection{The probabilistic time-varying vector autoregressive-1 process}
%%%%%%%%%%%%%%%%%%%%%%%%%%%%%%%%%%%%%%%%%%%%%%%%%%%%%%%%%%%%%%%%%%%%%%%%%%%%%%%%%%%%%%%%%%

We recently studied a probabilistic time-varying AR(1) process for detecting autocorrelation changes in univariate systems \cite{pTVAR}. Here we investigate a multivariate extension of this model, the time-varying probabilistic vector AR(1) model, tvPVAR(1); for simplicity, we refer to this method as \emph{ac/pooled} in the later comparisons. A non-probabilistic state space variant of this model was previously studied in \cite{Ives_Dakos_tvarss}.  Compared to the standard AR(1) process in Eq.~\ref{eqn:ar1_definition}, the time-varying model allows time-dependent variation in the model parameters. 

The tvPVAR(1) model is defined as the recursion

\begin{equation}
\label{eqn:mv_tvar_definition}
	\boldsymbol{X}_{t+1} =  \Phi_t \cdot \boldsymbol{X}_t  + \boldsymbol{\epsilon_t}
\end{equation} 

where \(\boldsymbol{X}_t\) is the \(D\)-dimensional state vector at time \(t\), \(\Phi_t\) the autoregressive matrix, \(\epsilon_t\) the multivariate Gaussian random variable with covariance matrix \(\Sigma\). The degrees of freedom  grow rapidly as a function of dimensionality, which makes parameter estimation challenging especially when the sample size is low compared to the dimensionality of the data. We are hence making certain simplifying assumptions. First, we assume that \(\Sigma\) is diagonal and constant over time. Second, we assume that \(\Phi_t = \phi_t I\), where \(\phi_t\) is a real number for all \(t\) and \(I\) is the identity matrix. The latter assumption amounts to parameter pooling (whence the name ac/pooled), which means that a single parameter ( \(\phi_t\) ) represents several units. Intuitively, this provides a measure for the average systemic autocorrelation. 

The probabilistic formulation requires us to define the likelihood of the data, and the priors for the model parameters. Likelihood for the data \(\boldsymbol{X_t}, \, t = 1, \ldots, T\) is given by

\begin{equation}
\label{eqn:tvpvar_likelihood}
	\mathcal{L}(\Phi_t, \Sigma | \boldsymbol{X}_t) = \prod_{t = 1}^{T-1} \text{MVN}(\boldsymbol{X}_{t+1} | \phi_t I \cdot \boldsymbol{X}_t, \Sigma),
\end{equation} 

where MVN refers to the multivariate normal distribution. 

Regarding the prior distribution, we use a Gaussian process (GP) prior for \(\phi_t\). A Gaussian process \(\mathcal{GP}(M, K) \) is formally defined as a collection of random variables where each finite collection of these variables is multivariate normally distributed with mean \(M\) and covariance \(K\) \cite{rasmussen_wilson}. GPs can be used for nonparametric regression. 

We utilize the Mat\`ern-$3/2$ covariance function that models the covariance between two random variables \(X_i\) and \(X_j\) as \(k_{3/2}(\rho, \alpha) = \alpha^2 \left( 1 +  \frac{\sqrt{3}r}{\rho} \right) \exp(-\frac{\sqrt{3}r}{\rho}),\) where \(\alpha^2\) is the process variance, \(\rho\) the length scale and \(r = |X_i - X_j|\) \cite{rasmussen_wilson, stein1999interpolation}. This covariance function restricts the posterior of \(\phi_t\) to be a continuous and differentiable function. This is a reasonable condition that allows flexibility in the model while avoiding overfitting to occassional large deviations. The length scale parameter \(\rho\) controls the dependence over time, while the process variance \(\alpha^2\) controls the average distance from the mean \(M\). We set \(M = 0 \) and \(\alpha = 1\), which restrains a majority of the prior values between -1 and 1. This is a justified choice as autoregressive-1 models are stationary if and only if the autoregressive parameter is within this interval.  Length scale \(\rho\) was set, unless otherwise noted, to the length of the time series. We used the Cholesky factored parameterization of GPs for posterior sampling \cite{kuss_rasmussen}. This models the process as a latent vector \(\eta\) which is mapped to the output space as \(\phi_t = L\eta\) where \(L\) is the lower triangular matrix with positive diagonal from the Cholesky decomposition \(k_{3/2} = LL^T\).

\begin{figure}[h!] % Data and inference example figure
\centering
\includegraphics[scale=0.075]{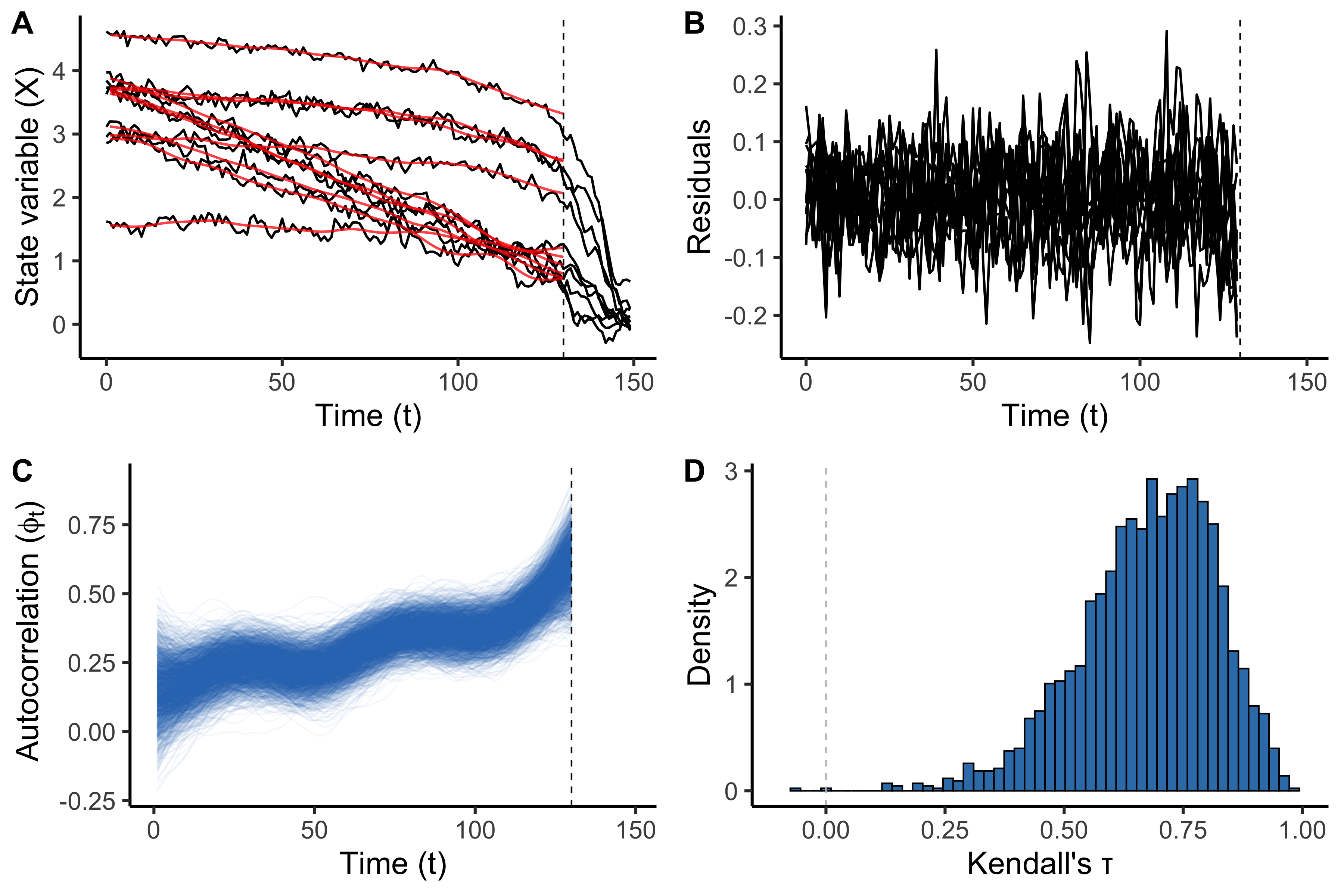}
\caption{Early warning signal detection with the proposed ac/pooled indicator (tvPVAR(1) model) in simulated ecological community data. \textbf{A} The observed system state \(X_i, i = 1, \ldots , N\) as a function of time (black curves). The system gradually approaches a tipping point before a system-wide collapse occurs starting at the vertical dashed line (\(T = 134\)). The red lines correspond to the estimated time series trends based on the Gaussian kernel smoother applied separately on the individual features. Detrending removes mean-level variations unrelated autocorrelation. \textbf{B} Residuals from the detrending process are used to look for the EWS.  \textbf{C} An increasing trend can be observed in the posterior samples of the autocorrelation parameter \(\phi_{t}\) from tvPVAR(1). \textbf{D}  Posterior of the Kendall's rank correlation for \(\phi_{t}\) verifies the observation that the average autocorrelation is increasing. More than 99.9 \% of the posterior mass is above 0 indicating strong posterior evidence for an increasing average autocorrelation and an EWS before the observed state transition.} 
\label{fig:fit_example}
\end{figure}

The fitting procedure is illustrated in Fig.~\ref{fig:fit_example}. Hypothesis testing was carried out by first computing Kendall's \(\tau\) for each posterior sample for \(\phi_t\). This provides a posterior distribution for \(\tau\), and the mass of this distribution on the positive side of the real line reflects the posterior probability of an increasing autocorrelation in the time series. The "Bayesian \emph{P}-value" \cite{gelman_BDA} can then be computed as the proportion on the negative side, facilitating comparison with non-probabilistic methods that generate frequentist P-values. We set the EWS detection level at \(P = 0.1\) and compare the methods in terms of the standard true positive rate (TPR) and true negative rate (TNR). 

We implemented the tvPVAR(1) model in the probabilistic programming language Stan \cite{stan}, utilizing the R interface RStan, and used its No-U-Turn variant of the Hamiltonian Monte Carlo algorithm with 2 chains, both with 2000 iterations for a given fit to sample the posterior. Sampling convergence was assessed with the \(\hat{R}\) statistic which remained below the recommended limit 1.1 \cite{gelman_rubin}. In addition, we encountered no divergent transitions indicating that the algorithm had converged and produced reliable estimates. 

%%%%%%%%%%%%%%%%%%%%%%%%%%%%%%%%%%%%%%%%%%%%%%%%%%%%%%%%%%%%%%%%%%%%%%%%%%%%%%%%%%%%%%%%%%
\subsection{Simulation model} \label{simulation_model}
%%%%%%%%%%%%%%%%%%%%%%%%%%%%%%%%%%%%%%%%%%%%%%%%%%%%%%%%%%%%%%%%%%%%%%%%%%%%%%%%%%%%%%%%%%

We evaluate the performance of the EWS detection methods based on simulations from a well-studied ecological model \cite{lever_PP}. The model characterizes systems with competition and mutualism, such as plant-pollinator interactions. 
The deterministic part of the model consists of logistic growth which is stimulated by intergroup mutualism and limited by competition within the same group. The model is defined by the stochastic differential equation

\begin{align}
\begin{split}
\label{eq:simulation_model}
dX_i^{(P)} &= X_i^{(P)} \left( r_i^{(P)}  + \frac{\Sigma_{k=1}^{S_P} \gamma_{ik}^{(P)} X_i^{(A)}}{1 + h^{(P)} \Sigma_{k=1}^{S_P} \gamma_{ik}^{(P)} X_i^{(A)}} - \Sigma_{j = 1}^{S_A}c_{ij}^{(P)} X_j^{(P)} \right)  dt + \sigma_{P, i}dW 
\end{split}
\end{align}

where \(X_i^{(P)}\) represents the abundance of pollinator species i, \(r\) the growth rate and \(h\) the half saturation constant, which was set to 0.5 for all species. The matrices \(\gamma_{ik}\) and \(c_{ij}\) represent the intergroup mutualism and interspecies competition, respectively. The last term is the stochastic part of the system, a Wiener process with variance \(\sigma_{P, i}^2\). The superscripts \(P\) and \(A\) refer to pollinators and plants, respectively. The equation above describes the dynamics of pollinator species \(i\), and the corresponding equation for plant species \(i\) can be recovered simply by interchanging the labels \(A\) and \(P\). 

The system can be pushed towards a critical transition by gradually decreasing the growth rate of the pollinator species \cite{lever_PP}, which could result from increasingly harsh environmental conditions, for instance. We randomly sampled initial pollinator growth rates \(r^{(P)}\) from \(\mathcal{N}(0, 0.1^2)\), and decreased them linearly to -1.5 during the simulation time, except in the cases where a group of pollinators were left undisturbed. In the latter case the growth rate was kept constant over the simulation. We randomly sampled rest of the parameters from the following distributions: \(r^{(A)}_i \sim \mathcal{N}(-0.1, 0.05^2)\); \(\gamma_{ij}^{(A)}, \gamma_{ij}^{(B)} \sim \text{Unif}(0.6, 1)\) for the off-diagonal elements and \(\gamma_{ij}^{(A)} = \gamma_{ij}^{(B)} = 1\) when  \(i = j\); \(c_{ij}^{(A)}, c_{ij}^{(B)} \sim \text{Unif}(0, 0.1)\) for the off-diagonal elements and \(c_{ij}^{(A)} = c_{ij}^{(B)} = 0.3\) when \(i = j\). We set the initial abundances to 2.5 for all species and then simulated the dynamics for 20 time points with constant conditions, during which the system settled into a stable state, and then discarded this settling period before EWS analysis. The stochastic noise \(\sigma\) was set to 0.1 in all cases. The chosen parameter sampling distributions and constants were based on previous studies \cite{weinans_2021, lever_PP}.

We used the Euler-Maruyama discretization with time-step \(\Delta t = 0.01\) to simulate the model and discarded all but every 100th observation, giving integer valued time points. For each replicate we required all species to be present after the settling period and defined a species to be present if its abundance is larger than 0.05. If this condition was not met, we repeated the parameter sampling process until a viable community emerged. We defined extinction to occur when any of the community species fell below 0.05. We only used the part preceding extinction in the EWS detection. 

To assess the performance and robustness of the various EWS indicators in different settings, we generated data sets with varying data characteristics. For the first part of the experiments we simulated a community with \(D = 10\) species in total, varied the number of perturbed pollinators from 1 to 5 and simulated \(T = 150\) time points per replicate with no observation error. In the second part we simulated data with \(D = 10\), \(T = 150\) and included random Gaussian observation error, with standard deviations 0, 0.05, 0.1 and 0.2. In the third part we used time series lengths of \(T = 50, 150, 250\), with \(D = 10\) and no observation error. In the final part we varied the total number of features \(D = 4, 10, 20\), with \(T = 150\) and no observation error. In order to assess the indicators' specificity, we also generated data with corresponding data characteristics but where the conditions were kept constant, and no extinction took place. In each distinct set of data characteristics 50 replicates were generated, and in every simulation half of the community species were plants and half were pollinators. 

All experiments and models were implemented in R.

%%%%%%%%%%%%%%%%%%%%%%%%%%%%%%%%%%%%%%%%%%%%%%%%%%%%%%%%
%%%%%%%%%%%%%%%%%%%%%%% RESULTS %%%%%%%%%%%%%%%%%%%%%%%%
%%%%%%%%%%%%%%%%%%%%%%%%%%%%%%%%%%%%%%%%%%%%%%%%%%%%%%%%
\section{Results}

%%%%%%%%%%%%%%%%%%%%%%%%%%%%%%%%%%%%%%%%%%%%%%%%%%%%%%%%
\subsection{Simulation benchmark}
%%%%%%%%%%%%%%%%%%%%%%%%%%%%%%%%%%%%%%%%%%%%%%%%%%%%%%%%

% (TPR; ANOVA \(P = 1.36\cdot 10^{-6}\); no significant difference in TNR)

In this section we compare the performance of the alternative EWS indicators presented in \emph{Methods} in ecological simulations. We limit our presentation here to the five top-performing autocorrelation-based methods, based on average TPR over all of our experiments. This filtering process excluded the degenerate fingerprinting and MAF eigenvalue indicators, and retained the ac/pooled, ac/mean, ac/max, ac/MAF and var/MAF.  

\begin{figure}[h!] % Benchmark figure
\centering
\includegraphics[scale=0.1]{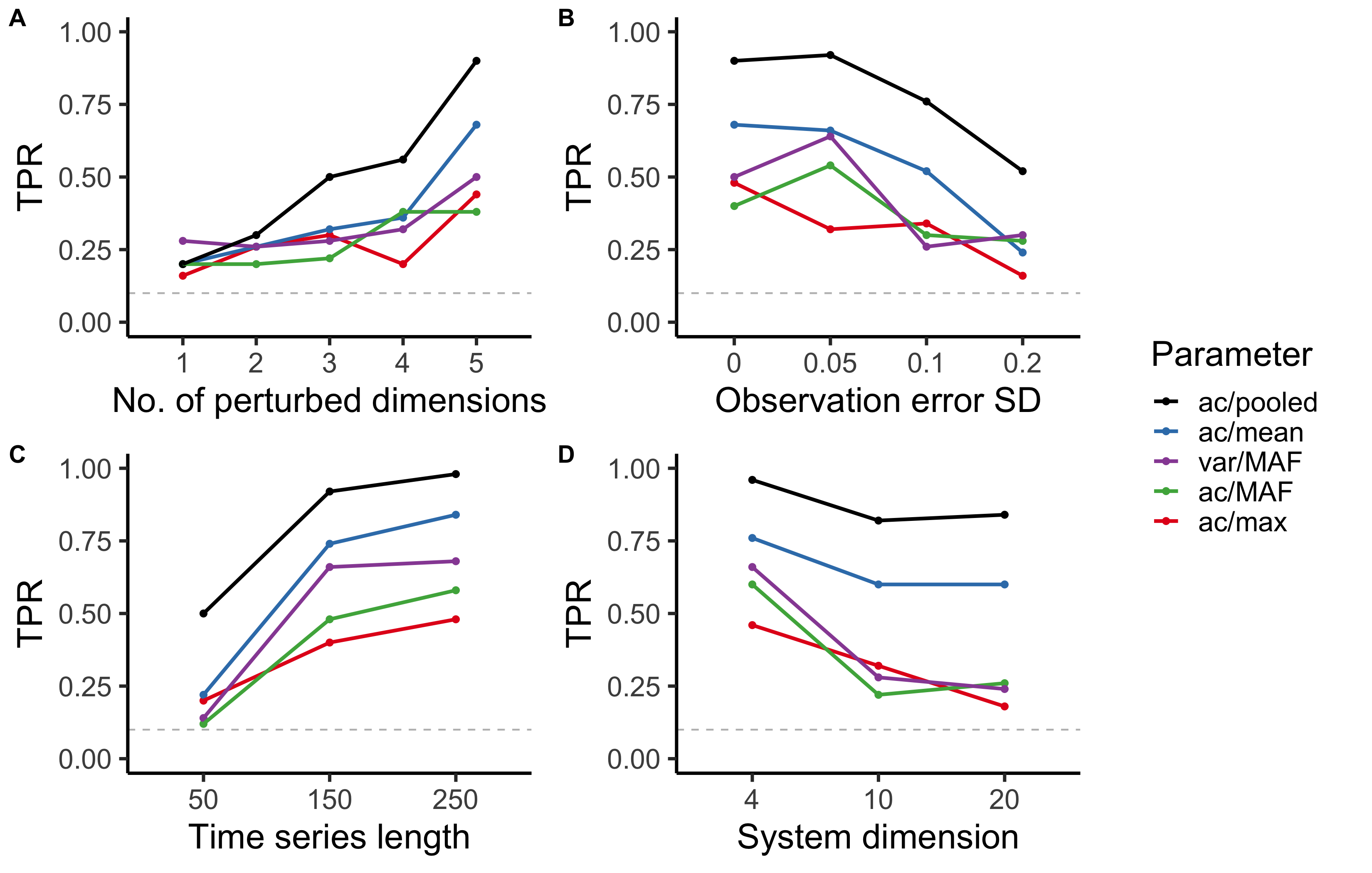}
\caption{EWS detection accuracy measured in true positive rate (TPR) depends on the data characteristics. \textbf{A} TPR increases as a function of perturbed dimensions. All replicates had 10 dimensions in total. \textbf{B} Increasing Gaussian additive observation error is associated with decreasing TPR. \textbf{C} Increasing time series lengths increase classification accuracy. The analyzed data sets were slightly shorter than indicated as only the part prior to the transition was used in analysis, and the exact number varied by simulation. The average time series lengths across the three groups were 49, 141 and 234, respectively. \textbf{D} The total dimensionality of the system also influences the TPR, when half of the features are perturbed. In all panels the horizontal dashed line marks the theoretical level of a random classification, 0.1. At each x-axis value in each panel the TPR is based on 50 replicates of the simulation with randomly selected parameter and initial values. We also tested the models' ability to correctly classify negative signals in data where conditions were kept constant. Results for this experiment have been omitted, as we did not observe significant differences in performance of the evaluated methods (ANOVA P = 0.097).} \label{fig:full_restults}
\end{figure}

We studied the effect of four different data characteristics on the detection performance: the number of perturbed features when the full dimensionality is kept constant (at 10), Gaussian additive observations error with four levels of standard deviations, time series length, and the total dimensionality of the system. Fig.~\ref{fig:full_restults} provides a graphical presentation of the results. 

The true positive rate (TPR) in EWS detection increased with the number of affected features (the number of different pollinator species). No clear differences between the alternative indicators were observed when only 1-2 features were affected. However, when a larger fraction of the the system, or a higher number of features was affected, a clear distinction between the methods emerges and ac/pooled achieves superior classification accuracy. 

Regarding observation error, the performance for all indicators decreased at the error level 0.1 or higher, compared to the noise-free case. At the lower error levels we observed mixed results, with increasing accuracy in some cases. Random variation may explain these differences (ANOVA \(F = 0.33\), \( P = 0.86\) between error levels 0 and 0.05). 

Increasing time series length led to a better accuracy. At \(T = 50\) ac/pooled achieved a TPR of approximately 0.5, whereas the other models performed only slightly above the theoretical level for random guess, 0.1. We observed a large further improvement in EWS detection with the longer \(T = 150\) set. Difference between 150 and 250 time points did not amount to a large improvement in TPR. This would either imply that the accuracy began to saturate, or that substantially larger amounts of samples are needed for further improvements in the TPR. We have omitted the analysis of longer time series in the present work because the longer time series are increasingly slow to model, and because our analysis is primary motivated by the practically important set of biomedical and ecological applications where the availability of longitudinal observations is limited to a few dozen time points.

By varying the total number of features in the data we observed, perhaps surprisingly, that EWS detection accuracy was best at the lowest-dimensional system \(D = 4\). The dimensionalities of \(D = 10\) and \(D = 20\) had reduced, and approximately similar performance. The better performance at the lowest dimension level might be explained by a lower level of mutualistic links, which in turn causes transients to be short lived. This could be detected visually from the time series, as the lower dimension cases experienced a more sudden collapse compared to the higher dimensional cases that collapsed in a more gradual fashion. 
% TODO: HARKITSE KUVAA LATER

In summary, our proposed indicator ac/pooled achieved the best performance (average TPR over all experiments 0.71), compared to ac/mean (0.51), var/MAF (0.4), ac/MAF (0.34), ac/max (0.31). One-way analysis of variance (ANOVA) indicated statistically significant differences between these methods (\emph{F} = 10.2, \(P = 1.4 \cdot 10^{-6}\)). We also looked at all of the aforementioned aspects in data with constant conditions and no EWS signal. Here, we found no meaningful differences between any of the methods in TNR (ANOVA \(F = 2.0\), \(P = 0.097\) over all experiments), which was close to the theoretical value expected to be get with a random guess, 0.9.  

%%%%%%%%%%%%%%%%%%%%%%%%%%%%%%%%%%%%%%%%%%%%%%%%%%%%%%%%
\subsection{Sensitivity analysis}
%%%%%%%%%%%%%%%%%%%%%%%%%%%%%%%%%%%%%%%%%%%%%%%%%%%%%%%%

The detected EWS signal depends on chosen hyperparameters, and the values can cause spurious false positives or false negatives. Here, we investigated the effect of data detrending bandwidth and sliding window length, or Gaussian process length scale prior for the probabilistic model, on the EWS detection in representative time series. 

\begin{figure}[h!] % Sensitivity analysis figure
\centering
\includegraphics[scale=0.075]{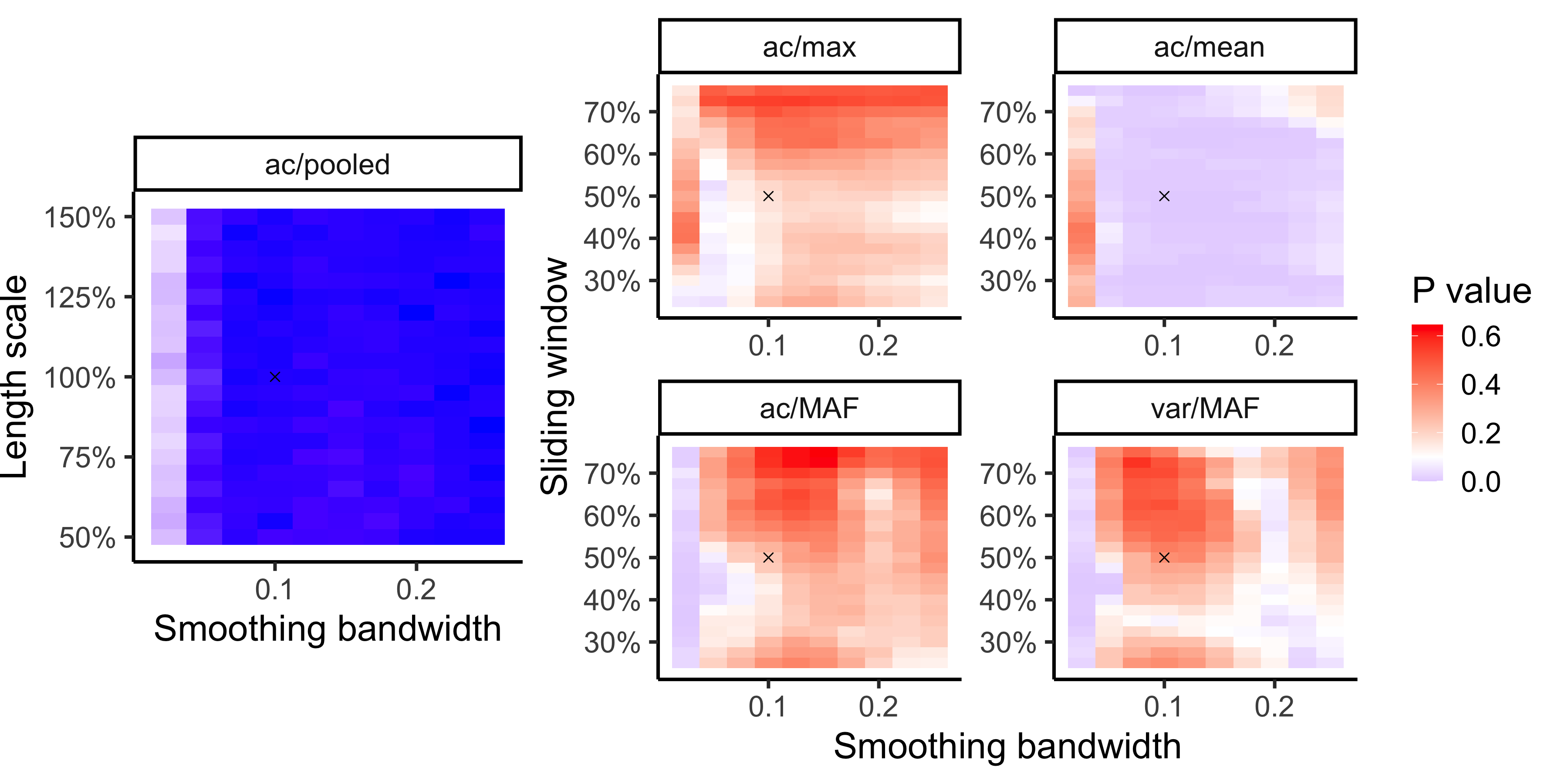}
\caption{Statistical evidence of the detected EWS depends on the hyperparameter combinations. The heatmap colors denote model evidence (posterior evidence and \emph{P}-value for ac/pooled and all other models, respectively) for an increasing systemic autocorrelation at the hyperparameter values. The y-axis values represent proportion of the time series length. The black cross in the panels denotes the hyperparameter values we used in other experiments, Gaussian smoothing bandwidth of 0.1, and 100\% and 50\% of the times series length for length scale and sliding window, respectively.} 
\label{fig:sensitivity}
\end{figure}

We noticed that for ac/pooled and ac/mean the choice of these parameters did not notably influence the EWS detectability, and ac/pooled in fact produced posterior evidence exceeding the EWS detection limit in all cases. At the lowest levels of the detrending bandwidth the \(P\)-values for both methods decreased, see Fig. \ref{fig:sensitivity}. For the other methods the experiment showed that an EWS was correctly identified only in a small set of hyperparameter combinations, indicating remarkable sensitivity to critical modeling choices. 

In time series with constant conditions and no expected EWS the results were more uniform: all methods correctly identified a true negative with practically all hyperparameter combinations (results not shown).

%%%%%%%%%%%%%%%%%%%%%%%%%%%%%%%%%%%%%%%%%%%%%%%%%%%%%%%%%%%
%%%%%%%%%%%%%%%%%%%%%%% DISCUSSION %%%%%%%%%%%%%%%%%%%%%%%%
%%%%%%%%%%%%%%%%%%%%%%%%%%%%%%%%%%%%%%%%%%%%%%%%%%%%%%%%%%%
\section{Discussion}

% Motivointi / kertaus

Early warnings have become an active area of research in the study of complex systems. Earlier attempts to design multivariate early warnings have generalized univariate indicators into multivariate context, as summarized in a recent review \cite{weinans_2021}. These methods quantify (auto)correlation and variance in multivariate systems by individual features, average values, or optimized subspaces, and rely on simplified representations of multivariate systems. Related approaches have been also proposed based on neural networks \cite{NeuralNetwork_EWS}, network analysis \cite{network_EWS}, and epidemic models \cite{covid_EWS}. These techniques typically rely on relatively large sample sizes and manual parameter adjustment, or lack explicit generative models for the data. These shortcomings form bottlenecks for practical application and interpretation. There is room for developing alternatives that are applicable to longitudinal data sets with limited sample sizes, provide explicit quantification of uncertainties, and avoid the need for manual parameter adjustment.

Our work is motivated by biomedical and ecological applications, where the number of available time points is typically low even in the best case scenarios. Therefore we have limited our experiments to relatively low sample sizes of up to 250 time points. It is noteworthy that the typical EWS methods in the literature generally rely on several hundreds or even thousands of time points, while these sample sizes are inaccessible in many applications. Hence, our experiments also provide a useful comparison for the alternative methods in the low sample size scenarios. Furthermore, the importance of analysing uncertainties are emphasized with smaller sample sizes. 

Our current work is an attempt to construct enhanced early warnings for multivariate time series. We have addressed the above-mentioned shortcomings by constructing a probabilistic variant of the time-varying vector autoregressive-1 model ("ac/pooled") that can detect early warnings of critical slowing down and resilience loss in multivariate systems. Whereas a similar pooling strategy could be considered for variance and other common EWS indicators, we have exclusively focused on autocorrelation-based methods in the current work because these outperformed the variance-based indicators in our initial experiments and have shown robustness also in other recent benchmarking studies \cite{weinans_2021, dakos_robustness_of_variance_and_autocorrelation}.

One of the advantages in using the probabilistic framework is that it makes the analysis of statistical certainty more straightforward. A posterior distribution for the test statistic, Kendall's \(\tau\) can directly be computed from the model parameter \(\phi_t\). In contrast, in the non-probabilistic setting one needs to resort to indirect and time-consuming surrogate data analysis methods to generate an approximate sampling distribution for the test statistic. This can be relatively simple when a data generating model is known as in our simulation experiments. However, the data generating processes are often unknown in practice and need to be approximated in order to generate surrogate models. In univariate setting, for instance, the ARMA(p, q) model has been used to identify the optimal ARMA model parameters, in order to generate surrogates from this model \cite{Dakos_2012}. In multivariate context the corresponding model is VARMA(p, q), but fitting this model is slow and potentially unreliable with higher dimension. Hence, the ability to directly estimate uncertainty in model parameters without such extra steps is beneficial. Another key advantage of the probabilistic framework is the ability to use prior distributions to regularize model fitting, and to incorporate available knowledge. This can be particularly useful when sample sizes are limited. By utilizing Gaussian process (GP) priors on \(\phi_t\) we could restrict its posterior to differentiable functions, and by GP hyperparameter selection we could emphasize longer term trends in the target variable which are of most interest in EWS context. On the other hand the Mat\'ern-3/2 covariance structure remains sufficiently flexible to detect relatively sudden changes as well \cite{stein1999interpolation, rasmussen_wilson}.

Benchmarking experiments showed systematic and robust improvements of the new model compared to available alternatives. As expected, the detection accuracy was in general better with larger perturbations. In all experiments, our proposed probabilistic multivariate indicator (ac/pooled) was systematically more sensitive than the other alternative autocorrelation based models and shorter time series were sufficient to achieve similar levels sensitivity than with alternative indicators. The highest TPR was achieved in systems with the smallest dimensionality. We speculate that such behavior could occur for instance when transients are more short-living with less species and mutualistic links that stabilize the system. All models performed equally well in terms of the true negative rate, close to their theoretical true negative rate corresponding to a random guess. No significant trends favouring any particular model in this regard were observed. Finally, our proposed method (ac/pooled) also outperformed the other methods in hyperparameter sensitivity analysis as it correctly detected the true EWS in a representative time series at all hyperparameter combinations.

% Limitations and potential future extensions of the work

The current work provides a proof-of-concept study on the potential of probabilistic multivariate early warning signals based on a single well-characterized ecological model that has been used also in other EWS studies. Additional simulation models, real case studies, and variations in data resolution, interaction structures, multiplicative noise, or large dimensionality, will help to assess the broader utility of the approach in practical scenarios \cite{Clements_2015, arkilanian2020}. 

The proposed method relies on a simple diagonal structure for the transition matrix with tied parameters. This aggregates information across the system and reduces the deterministic part of the dynamics into a single variable. The downside is that this will neglect interactions and does not inform us about the specific parts of the system that are affected and under risk. Future extensions could hence benefit from allowing off-diagonal terms in the transition matrix with a suitable regularization or sparsity inducing priors in order to increase model flexibility and capture important additional aspects of covariance within the system. A similar but more restricted approach would be to enhance automated feature selection by allowing the diagonal elements to vary independently, analogously to the maximal lag-1 autocorrelation in the non-probabilistic context. This could be regularized for instance with a composite GP prior, consisting of a common process, as in our model, in addition to separate GPs for the distinct elements which would be used to characterize additional, individual trajectories. 

Furthermore, while quantifying uncertainties in parameter inference, our current method lacks an explicit model for observation error; adding this would allow the analysis of alternative error structures and potentially expand the scope of the method to different types of systems. Whereas a time-varying vector autoregressive-1 state space model has been studied in EWS context as a potential solution \cite{Ives_Dakos_tvarss}, this was only tested on 2-dimensional simulated data and convergence issues could arise in higher dimensions, and parameter estimation with state space models can be challenging even in the simplest cases \cite{state_space_models_review}.
Further extensions could consider variance, network structure, and other aspects of the system. For instance, principal component analysis has been used to detect changes in maximal variance and to identify features that are potentially most vulnerable \cite{dakos_2018_multispecies, chen_eigenvalues}. Probabilistic PCA \cite{PPCA} could add sensitivity to these analyses by explicitly distinguishing between measurement error and random variations in the data. Incorporating other aspects of dynamics, such as estimated exit times \cite{arani_2021} or memory properties \cite{Khalighi2022}, could provide further means to enhance EWS, and combining the analysis of longitudinal time series and aspects of multivariate survival analysis (see e.g. \cite{salosensaari2021}) as prior information, could provide interesting avenues for future research.

% Lopetus

The detection of early warning signals for critical transitions is a highly topical yet challenging task. Given the limitations typically encountered in applied scenarios, it is paramount that the available information can be utilized in the most optimal way, and uncertainties communicated effectively. Our proof-of-concept study presents a step towards this direction, providing an example on how probabilistic multivariate analysis could provide the means to develop more sensitive, robust, and intuitive alternatives for the currently available early warnings signals in complex dynamical systems.

\section{Acknowledgements}
This work has been supported by Academy of Finland (decisions 295741, 330887) and by Turku university graduate school (UTUGS). The authors wish to acknowledge CSC – IT Center for Science, Finland, for computational resources. The authors declare no conflict of interest.

\section{Code availability}

Source code for the experiments is available at 10.5281/zenodo.6472720

\bibliographystyle{splncs04}
\bibliography{bibliography}

\end{document}